\documentclass[aps,reprint,preprintnumbers]{revtex4-1}
\usepackage{bm}
\usepackage{amsmath}
\usepackage{graphicx}

\begin{document}
\title{Inclusive semi-leptonic decays from lattice QCD}
\date{\today}

\author{Paolo \surname{Gambino}}
\affiliation{Dipartimento di Fisica, Universit\`a di Torino and
  INFN Torino\\ Via P. Giuria 1, I-10125, Torino, Italy}
\author{Shoji \surname{Hashimoto}}
\affiliation{Theory Center, Institute of Particle and Nuclear Studies,
  High Energy Accelerator Research Organization (KEK), Tsukuba
  305-0801, Japan}
\affiliation{School of High Energy Accelerator Science,
  The Graduate University for Advanced Studies (SOKENDAI),
  Tsukuba 305-0801, Japan}

\begin{abstract}
  We develop a method to compute inclusive semi-leptonic decay rate of
  hadrons fully non-perturbatively using lattice QCD simulations.
  The sum over all possible final states is achieved by a calculation
  of the forward-scattering matrix elements on the lattice,
  and the phase-space integral is evaluated using their dependence on
  the time separation between two inserted currents.
  We perform a pilot lattice computation for the
  $\bar B_s\to X_c\ell\bar{\nu}$ decay with an unphysical bottom quark
  mass and compare the results with the corresponding OPE calculation.
  The method to treat the inclusive processes on the lattice can be
  applied to other processes, such as the lepton-nucleon inelastic
  scattering.
\end{abstract}

\preprint{KEK-CP-376}
\maketitle

Quark-hadron duality plays a key role in perturbative
Quantum Chromodynamics (QCD) calculations of physical processes.
It states that hadronic processes can be calculated taking quarks and
gluons as final states, 
even though the actually observed final states are composed of
hadrons.
In order that the duality is satisfied,
the processes must be summed or smeared over all possible hadronic
final states in some kinematical range \cite{Poggio:1975af},
such as a region of invariant mass squared,
but it is not  {\it a priori} known how large the smearing should be.
A systematic approach to duality is based on 
the Operator Product Expansion (OPE) \cite{Shifman:2000jv,Bigi:2001ys},
which is constructed in the Euclidean domain and analytically
continued to the Minkowski domain.
In the context of heavy quark decays the OPE is an expansion in
inverse powers of the heavy quark mass,
or more precisely of the energy release,
and the analytic continuation entails an inevitable violation of duality.
While there are indications that duality violation  plays a
minor role in the analysis of inclusive semi-leptonic $B$ meson decays
to determine the Cabibbo-Kobayashi-Maskawa (CKM) matrix element $|V_{cb}|$
\cite{Alberti:2014yda,Gambino:2016jkc},
full control of the systematic error can only be achieved by
non-perturbative methods.
The current tension between the inclusive and  exclusive
determinations of $|V_{cb}|$ \cite{Tanabashi:2018oca,Gambino:2019sif}
makes any contribution in this direction timely.

Lattice QCD simulation provides a means of non-perturbative
QCD computation for various hadronic processes including heavy quark
decays. 
It has been successfully applied to the calculation of exclusive decay
form factors, which are essential for a precise determination of the CKM
elements, 
{\it e.g.} $K\to\pi\ell\bar{\nu}$, $B\to D^{(*)}\ell\bar{\nu}$,
{\it etc.} (see \cite{Aoki:2019cca} for their world averages),
while the study of inclusive processes is scarce, except for recent 
attempts to formulate methods 
to introduce an analytic continuation
\cite{Hashimoto:2017wqo} or a smearing \cite{Hansen:2017mnd}.
The inclusive processes are, on the other hand, difficult to treat on
the lattice because they consist of many physical states often
including multiple hadrons. 
To identify each amplitude and to sum over the phase space is 
nearly impossible due to the number of states involved. 
One may instead use analyticity and the optical theorem to relate the
total rate to another quantity that is calculable on the lattice.
This approach has been followed in simple cases, such as
the $e^+e^-\to q\bar{q}$ processes
\cite{Bernecker:2011gh,Feng:2013xsa,Francis:2013fzp,Lehner:2020crt}
and $\tau$-lepton decays \cite{Tomii:2017cbt,Boyle:2018ilm}, while the
application to $B$ meson semi-leptonic decays is much more
complicated \cite{Hashimoto:2017wqo}.

In this work we develop a novel and  general method to compute
the inclusive semi-leptonic decay rate on the lattice.
The method is based on a technique to calculate smeared spectral
density of hadron correlators \cite{Bailas:2020qmv}
(see also \cite{Hansen:2017mnd,Bulava:2019kbi} for a slightly
different strategy).
The extraction of the spectral density $\rho(\omega)$ of hadronic
correlation functions remains intractable, but once $\rho(\omega)$
 is smeared over some energy range, one can construct a good
approximation using the correlation functions calculated on the
lattice.
In semi-leptonic decays of hadrons, the phase-space integral
plays the role of this smearing. 
The method is systematically improvable as more
computational resources are made available.

In this paper we use the inclusive semi-leptonic 
decays 
$\bar B_s\to X_c\ell\bar{\nu}$ 
to demonstrate how the
method works.
Here, $X_c$ stands for all possible charmed states which may occur
with the quark-level decay process $b\to c\ell\bar{\nu}$.
After describing the kinematics of the decay and the method to
calculate the inclusive decay rate, we present a pilot lattice study. 

For the analysis of the $\bar B_s\to X_c\ell\bar{\nu}$ decay,
we assign a momentum $p^\mu$ to the initial $\bar B_s$ meson, and 
momenta $p_\ell^\mu$ and $p_{\bar{\nu}}^\mu$ to the leptons
$\ell$ and $\bar{\nu}$ in the final state, respectively.
Thus, the hadronic state $X_c$ has momentum $r^\mu=(p-q)^\mu$ with $q^\mu=(p_\ell+p_{\bar{\nu}})^\mu$.
The differential decay rate is written as
\cite{Manohar:1993qn,Blok:1993va}
\begin{equation}
  \label{eq:differential}
  \frac{d\Gamma}{dq^2dq^0dE_\ell}=
  \frac{G_F^2|V_{cb}|^2}{8\pi^3} L_{\mu\nu}W^{\mu\nu},
\end{equation}
where $G_F$ is the Fermi constant. 
The momentum transfer $q^\mu$ and the lepton energy $E_\ell$ are
evaluated in the rest frame of the initial $\bar B_s$ meson.
The leptonic tensor $L_{\mu\nu}$ is explicitly written as
$L^{\mu\nu}=p_\ell^\mu p_{\bar{\nu}}^\nu
-p_\ell \cdot p_{\bar{\nu}}g^{\mu\nu}
+p_\ell^\nu p_{\bar{\nu}}^\mu
-i\epsilon^{\mu\alpha\nu\beta}p_{\ell,\alpha}p_{\bar{\nu},\beta}$
for massless neutrinos.
The hadronic tensor $W^{\mu\nu}(p,q)$ is defined through
\begin{eqnarray}
  \lefteqn{
  W^{\mu\nu}(p,q) = \sum_{X_c}(2\pi)^3\delta^{(4)}(p-q-r)
  }
  \nonumber\\
  && \mbox{}
  \times\frac{1}{2E_{B_s}}
  \langle \bar B_s(\bm{p})|J^{\mu\dagger}|X_c(\bm{r})\rangle
  \langle X_c(\bm{r})|J^\nu|\bar B_s(\bm{p})\rangle.
\end{eqnarray}
It is summed over all possible final states $X_c$ to represent the
inclusive decay.
The electroweak current relevant for this decay mode is
$J^\mu=(V-A)^\mu=\bar{c}\gamma^\mu(1-\gamma_5)b$.

One can perform an integral over the lepton energy $E_\ell$ in
(\ref{eq:differential}), and the remaining integrals over $q^2$ and
$q^0$ can be rewritten in terms of $\omega$ and $\bm{q}^2$, energy
and spatial momentum squared of the final hadrons $X_c$,
respectively. 
Thus, the total decay rate can be calculated as
\begin{equation}
  \Gamma=\frac{G_F^2|V_{cb}|^2}{24\pi^3}
  \int_0^{\bm{q}^2_{\mathrm{max}}} d\bm{q}^2 \sqrt{\bm{q}^2}
  \sum_{l=0}^2 \bar{X}^{(l)},
  \label{eq:q2integ}
\end{equation}
where $\bm{q}^2_{\mathrm{max}}=((m_{B_s}^2-m_{D_s}^2)/2m_{B_s})^2$
and 
\begin{equation}
  \bar{X}^{(l)} \equiv
  \int_{\sqrt{m_{D_s}^2+\bm{q}^2}}^{m_{B_s}-\sqrt{\bm{q}^2}}
  d\omega\, X^{(l)}
  \label{eq:omega_integ}
\end{equation}
with
\begin{eqnarray}
  X^{(0)} & = & \bm{q}^2 (W^{00}-2W^{ii}), \label{eq:X0}\\
  X^{(1)} & = & -(m_{B_s}-\omega)q_k(W^{0k}+W^{k0}), \label{eq:X1}\\
  X^{(2)} & = & (m_{B_s}-\omega)^2(W^{kk}+2W^{ii}). \label{eq:X2}
\end{eqnarray}
Here, we take the momentum $\bm{q}$ in the $k$-th direction,
while the $i$-th direction is assumed to be perpendicular to that.
The repeated indices in (\ref{eq:X0})--(\ref{eq:X2}) are not
summed. 
The integral with respect to $\omega$ in (\ref{eq:omega_integ})
represents the sum over states that could appear for a given momentum
$\bm{q}$. 

On the lattice, as a counterpart of the hadronic tensor $W_{\mu\nu}$,
one can calculate the forward-scattering matrix elements of the form
\cite{Hashimoto:2017wqo}
\begin{equation}
C_{\mu\nu}^{JJ}(t;\bm{q})=\!\sum_{\bm{x}} 
        \frac{e^{i\bm{q}\cdot\bm{x}}}{2m_{B_s}}\langle \bar B_s(\bm{0})|
        J_\mu^\dagger(\bm{x},\!t)J_\nu(\bm{0},\!0)
        |\bar B_s(\bm{0})\rangle
        \label{eq:forward-scatteringME}
\end{equation}
from four-point functions 
including the interpolating operators for
the $\bar B_s$ meson state $|\bar B_s(\bm{0})\rangle$. 
Now we introduce the transfer matrix on the lattice $e^{-\hat{H}t}$
to express the time dependence of the matrix element in
(\ref{eq:forward-scatteringME}) as
\begin{equation}
  \frac{1}{V} \frac{1}{2m_{B_s}}
  \langle \bar B_s(\bm{0})|\tilde{J}_\mu^\dagger(-\bm{q})
  e^{-\hat{H}t}\tilde{J}_\nu(\bm{q})|\bar B_s(\bm{0})\rangle,
  \label{eq:trasfer-matrixME}
\end{equation}
where $\tilde{J}_\nu(\bm{q})$ denotes a Fourier transform of the
inserted current: $\tilde{J}_\nu(\bm{q})=\sum_{\bm{x}}e^{i\bm{q}\cdot\bm{x}}J_\nu(x)$.
On the other hand, the integral over $\omega$ in
(\ref{eq:omega_integ}) can be rewritten in the form
\begin{align}
  \int_0^\infty\!d\omega\,K(\omega,\bm{q})
  \langle \bar B_s(\bm{0})|\tilde{J}_\mu^\dagger(-\bm{q})
  \delta(\hat{H}-\omega)
  \tilde{J}_\nu(\bm{q})|\bar B_s(\bm{0})\rangle
  \;\;\;& \nonumber\\
  = 
  \langle \bar B_s(\bm{0})|\tilde{J}_\mu^\dagger(-\bm{q})
  K(\hat{H},\bm{q})
  \tilde{J}_\nu(\bm{q})|\bar B_s(\bm{0})\rangle.
  \label{eq:integral_with_kernel}
\end{align}
Here $K(\omega,\bm{q})$ represents an integral kernel determined by
the explicit form of the integrands (\ref{eq:X0})--(\ref{eq:X2}).
The $\omega$-integral is implicit on the right hand side;
all the intermediate states may exist between the currents.
Comparing the right hand side with
(\ref{eq:trasfer-matrixME}),
we find that the integral (\ref{eq:integral_with_kernel}) can be
evaluated 
if the kernel operator is well approximated by a polynomial of the
form
\begin{equation}
  K(\hat{H},\bm{q}) = k_0(\bm{q}) + k_1(\bm{q}) e^{-\hat{H}} + \cdots
  + k_N(\bm{q}) e^{-N\hat{H}}
  \label{eq:poly}
\end{equation}
with some coefficients $k_j(\bm{q})$, since the
matrix elements of the individual term on the right hand side are
nothing but $C_{\mu\nu}^{JJ}(t;\bm{q})$'s.

The best approximation of $K(\hat{H},\bm{q})$ can be achieved using
the Chebyshev polynomials.
We define a state
$|\psi_\mu(\bm{q})\rangle$ on which the kernel operator is
evaluated as
$|\psi_\mu(\bm{q})\rangle=e^{-\hat{H}t_0}
\tilde{J}_\mu(\bm{q})|\bar B_s(\bm{0})\rangle$.
A small time evolution $e^{-\hat{H}t_0}$ with a constant time $t_0$ is
introduced to avoid any potential divergence in 
$\langle\psi_\mu(\bm{q})|\psi_\nu(\bm{q})\rangle$.
We can then construct an approximation as
\begin{equation}
  \frac{\langle\psi_\mu|K(\hat{H})|\psi_\nu\rangle}{
    \langle\psi_\mu|\psi_\nu\rangle} \simeq
  \frac{c_0^*}{2} + \sum_{j=1}^N c_j^*
  \frac{\langle\psi_\mu|T_j^*(e^{-\hat{H}})|\psi_\nu\rangle}{
    \langle\psi_\mu|\psi_\nu\rangle}.
  \label{eq:Chebyshev_expansion}
\end{equation}
(The dependence on $\bm{q}$ is omitted for simplicity.)
$T_j^*(x)$ stands for the shifted Chebyshev polynomials, which are
derived from the standard Chebyshev polynomials $T_j(x)$ as 
$T_j^*(x)\equiv T_j(2x-1)$, so that they are defined in the range
$0\le x\le 1$.
Their first few terms are
$T_0^*(x)=1$, $T_1^*(x)=2x-1$, $T_2^*(x)=8x^2-8x+1$, and the others can be
obtained recursively by
$T_{j+1}^*(x)=(4x-2)T_j^*(x)-T_{j-1}^*(x)$.
Each term of 
$\langle\psi_\mu|T_j^*(e^{-\hat{H}})|\psi_\nu\rangle
/\langle\psi_\mu|\psi_\nu\rangle$
can be constructed from
$C_{\mu\nu}^{JJ}(t+2t_0)/C_{\mu\nu}^{JJ}(2t_0)
=\langle\psi_\mu|e^{-\hat{H}t}|\psi_\nu\rangle
/\langle\psi_\mu|\psi_\nu\rangle$.

The coefficients $c_j^*$ in (\ref{eq:Chebyshev_expansion})
are obtained from 
\begin{equation}
  c_j^*=\frac{2}{\pi}\int_0^\pi\!d\theta\,
  K\left(-\ln\frac{1+\cos\theta}{2}\right)\cos(j\theta),
\end{equation}
according to the general formula of the Chebyshev approximation.
The Chebyshev approximation is the {\it best} in the sense that its
maximum deviation in $x\in [0,1]$ is minimized among polynomials of
order $N$.

\begin{figure}[tb]
  \centering
  \includegraphics[width=7cm]{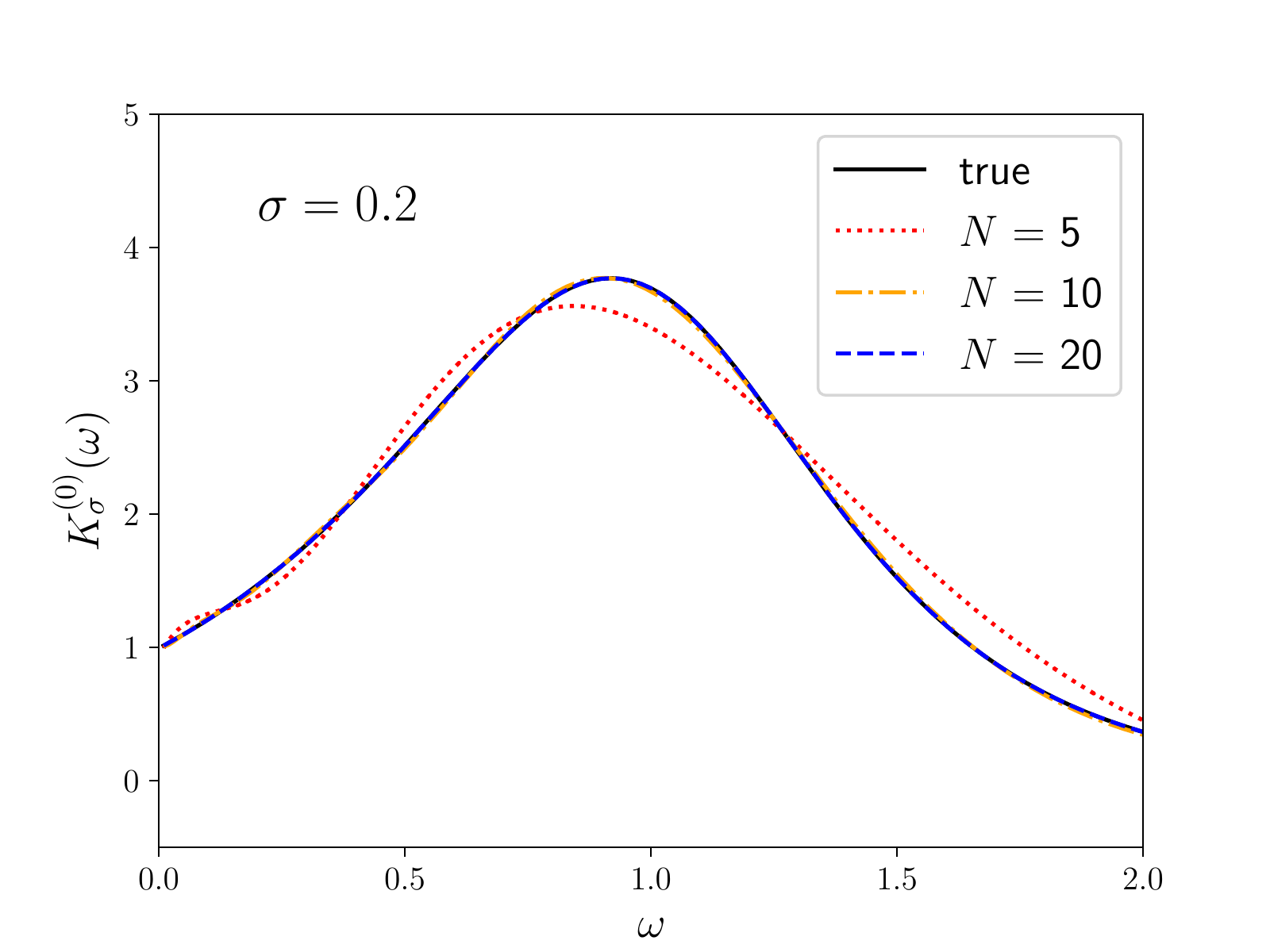}
  \includegraphics[width=7cm]{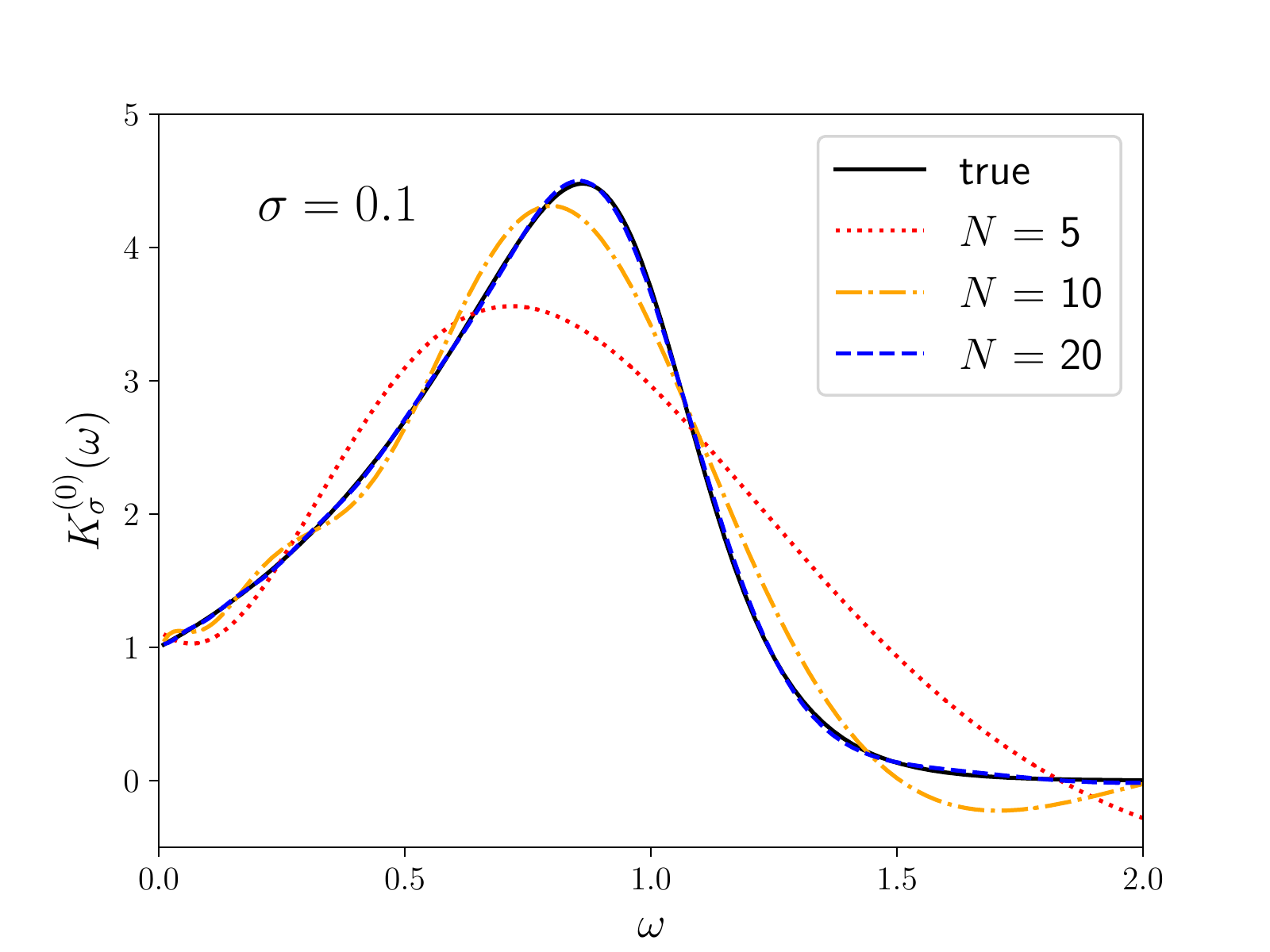}
  \includegraphics[width=7cm]{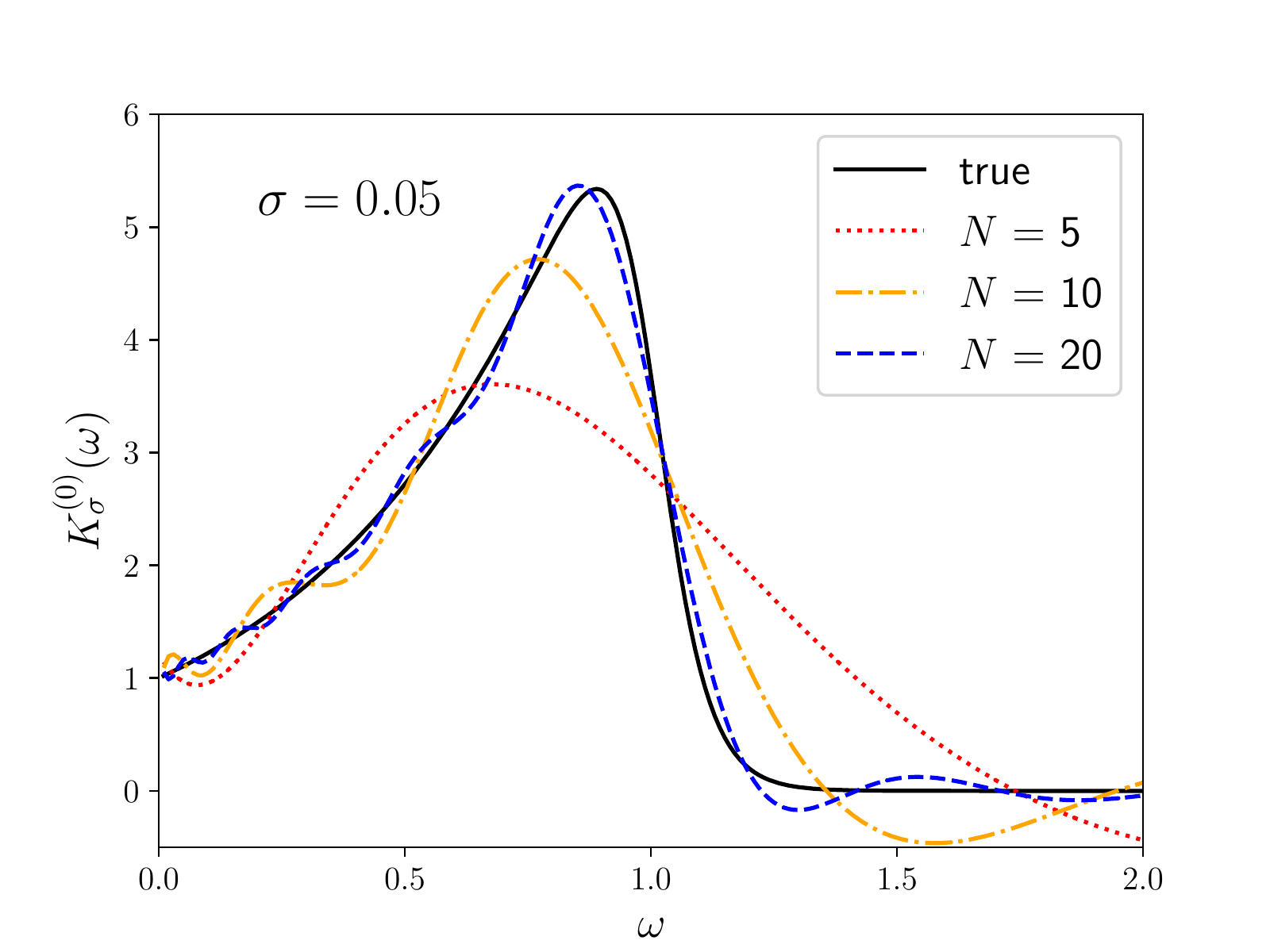}
  \caption{Approximation of the weight function
    $K_\sigma^{(l=0)}(\omega)$ with the Chebyshev polynomials of
    $e^{-\omega}$.
    For each value of the smearing width $\sigma$ (= 0.2 (top), 0.1
    (middle), 0.05 (bottom)), the approximations with the polynomial
    order $N$ = 5 (dotted), 10 (dot-dashed), 20 (dashed) are plotted
    as well as the true curve (solid curve).
  }
  \label{fig:weight_l=0}
\end{figure}

The integral kernel $K(\omega,\bm{q})$ is chosen as
\begin{eqnarray}
  K^{(l)}_\sigma(\omega)
  & = & e^{2\omega t_0} (-\sqrt{\bm{q}^2})^{2-l}(m_{B_s}-\omega)^l
              \nonumber\\
  && \times \theta_\sigma(m_{B_s}-\sqrt{\bm{q}^2}-\omega)
     \label{eq:kernel}
\end{eqnarray}
for $l$ = 0, 1, or 2 corresponding to $X^{(l)}$,
(\ref{eq:X0})--(\ref{eq:X2}).
An approximate Heaviside step function $\theta_\sigma(x)$ is
introduced to realize the upper limit of the $\omega$-integral.
In order to stabilize the Chebyshev approximation, 
we smear the step function over a small width $\sigma$.
For an explicit form, we chose
$\theta_\sigma(x)=1/(1+\exp(-x/\sigma))$.
The extra factor $e^{2\omega t_0}$ in (\ref{eq:kernel}) cancels the
short time evolution
$e^{-\hat{H}t_0}$ in $|\psi_\mu(\bm{q})\rangle$.

Fig.~\ref{fig:weight_l=0} demonstrates how
well $K_\sigma^{(l)}(\omega)$ is approximated with certain
orders of the polynomials, {\it i.e.} $N$ = 5, 10 and 20.
An example for $l=0$ is shown.
Here we take three representative values of $\sigma$:
0.2, 0.1 and 0.05 in  lattice units.
The comparison is made for parameters that roughly correspond to our
lattice setup: the inverse lattice spacing
$1/a\simeq$ 3.61~GeV, $am_{B_s}\simeq 1.0$, $t_0/a=1$.
The momentum insertion $\bm{q}$ is set to zero.
The kernel function is well approximated with relatively low orders of
the polynomials, such as $N=10$, when sufficiently smeared,
{\it e.g.} $\sigma$ = 0.2.
For smaller $\sigma$'s, the function exhibits a more rapid change near
the threshold $\omega=1.0$, and 
one needs higher orders, like $N=20$.
Eventually we have to take the limit  $\sigma\to 0$, and the error
due to finite $N$ has to be estimated.
For  $l=1$ and 2 the polynomial approximations are
better than those for $l=0$.

We perform a pilot study of the method described above using lattice
data computed on an ensemble with 2+1 flavors of M\"obius domain-wall
fermions
(the ensemble ``M-$ud$3-$s$a'' in \cite{Nakayama:2016atf}, which has
$1/a$ = 3.610(9)~GeV).
For the charm and bottom quarks in the valence sector, the 
same lattice formulation is used.
The charm quark mass $m_c$ is tuned to its physical value and
the $D_s$ and $D_s^*$ meson masses are 1.98 and 2.12~GeV,
respectively.
The bottom quark mass is taken as $2.44m_c$, which is substantially
smaller than the physical $b$ quark mass.
The corresponding $B_s$ meson mass is 3.45~GeV.
In this setup, the maximum possible spatial momentum in the
$B_s\to D_s\ell\bar{\nu}$ decay is
$(m_{B_s}^2-m_{D_s}^2)/2m_{B_s}\simeq$ 1.16~GeV.
The lattice volume is $L^3\times L_t=48^3\times 96$, and we calculate
the forward-scattering matrix elements with spatial momenta $\bm{q}$ of
(0,0,0), (0,0,1), (0,0,2) and (0,0,3) in units of $2\pi/La$.
The number of lattice configurations averaged is 100, and the
measurement is performed with four different source time-slices.

For a fixed spatial momentum $\bm{q}$, we compute a four-point
function to extract $C_{\mu\nu}^{JJ}(t;\bm{q})$ (more details of the lattice calculation are presented in
\cite{Hashimoto:2017wqo}).
We perform the $\omega$-integral (\ref{eq:omega_integ}) using the
representation (\ref{eq:Chebyshev_expansion}).
Matrix elements of the shifted Chebyshev polynomials are obtained
from $C_{\mu\nu}^{JJ}(t+2t_0;\bm{q})/C_{\mu\nu}^{JJ}(2t_0;\bm{q})$ at
various $t$'s (and $t_0=1$) by a fit with constraints
$|\langle\psi_\mu|T_j^*(e^{-\hat{H}})|\psi_\nu\rangle
/\langle\psi_\mu|\psi_\nu\rangle|<1$,
which is a necessary condition for the Chebyshev polynomials.

\begin{figure}[tb]
  \centering
  \includegraphics[width=8cm]{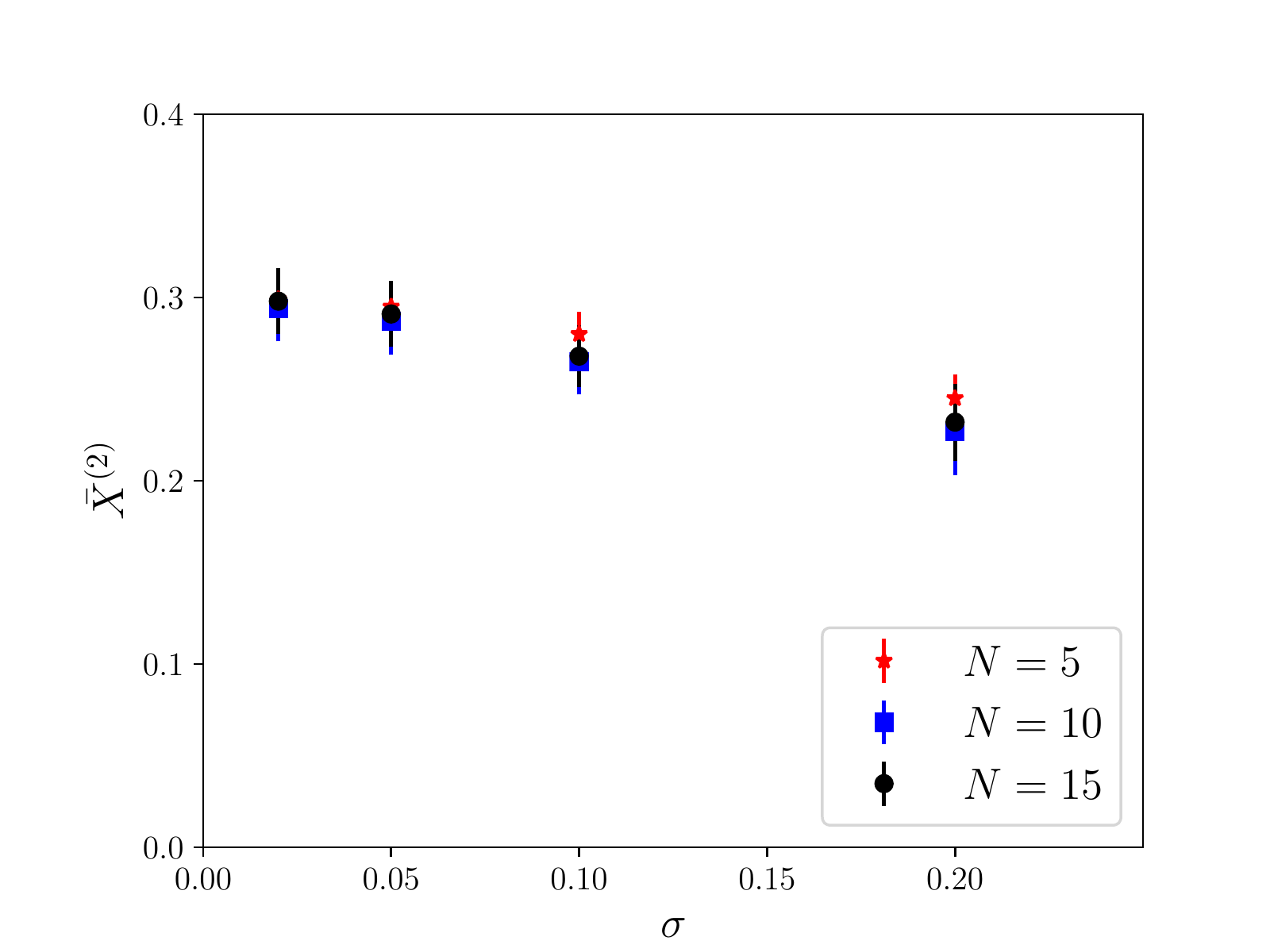}
  \caption{
    $\bar{X}^{(2)}$ at $\bm{q}=2\pi/La (0,0,1)$
    plotted as a function of the smearing width $\sigma$.
    Results with polynomial orders $N$ = 5, 10, 15 are shown.
  }
  \label{fig:X2}
\end{figure}

First, we inspect how well the Chebyshev approximation works
by comparing the results for $\bar{X}^{(2)}$ obtained with the
polynomial order $N$ = 5, 10, 15 at various values of $\sigma$,
the width of the smearing.
Fig.~\ref{fig:X2} shows that the dependence on $\sigma$ is mild and
the limit of $\sigma=0$ is already reached at around $\sigma=0.05$.
The dependence on $N$ is not significant, which indicates that the
approximation is already saturated at $N\simeq 10$.
This is crucial because the error of the lattice data is too large to 
constrain the matrix elements
$\langle\psi_\mu|T_j^*(e^{-\hat{H}})|\psi_\nu\rangle
/\langle\psi_\mu|\psi_\nu\rangle$ at $j\simeq$ 10 or larger.
The results for $\bar{X}^{(0)}$ and $\bar{X}^{(1)}$ show the similar
tendency.
We take $\sigma=0.05$ in the following analysis;
the results are within statistical error even if we extrapolate to
$\sigma=0$.

\begin{figure}[tbp]
  \centering
  \includegraphics[width=8cm]{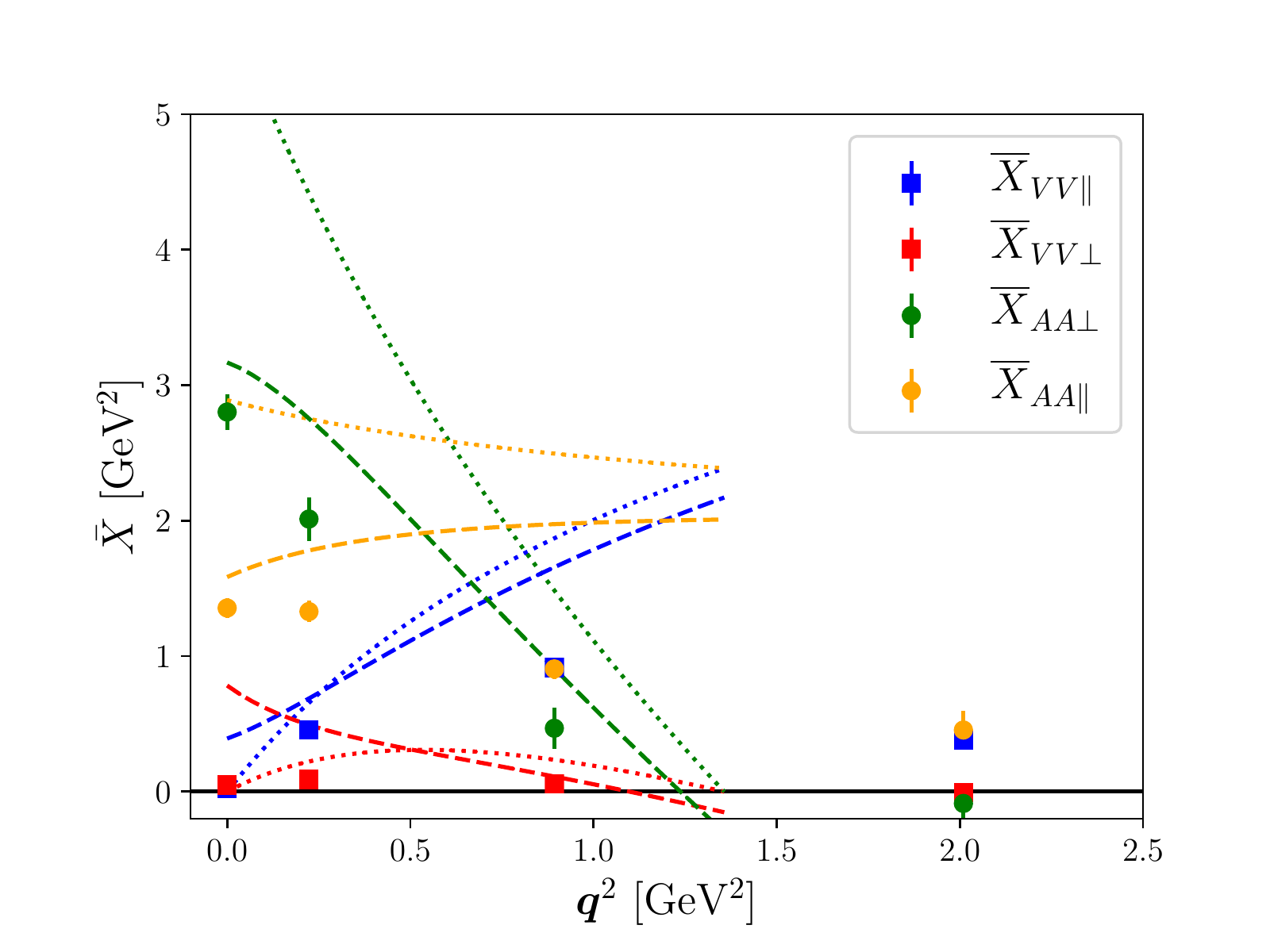}\\
  \caption{
    $\bar{X}$ as a function of $\bm{q}^2$ plotted
    in the physical unit. 
    Longitudinal ($\parallel$) and perpendicular ($\perp$)
    polarizations are plotted for vector ($VV$) and axial-vector
    ($AA$) channels.
    Dotted and dashed curves show the lowest order and $O(1/m^2)$ OPE estimates for
    each channel of corresponding color, respectively.
  }
  \label{fig:channel_q2}
\end{figure}

The lattice results for
$\bar{X}=\sum_{l=0}^2\bar{X}^{(l)}$
are compared with the OPE predictions in Fig.~\ref{fig:channel_q2}
as a function of $\bm{q}^2$.
Here, the results for different polarizations,
{\it i.e.}
longitudinal ($\parallel$: $\mu$, $\nu$ = 0 and 3) and
perpendicular ($\perp$: $\mu$, $\nu$ = 1 and 2)
directions to $\bm{q}$,
are separately plotted for
vector ($VV$, squares) and axial-vector ($AA$, circles) current
contributions.
The lowest order  and $O(1/m^2)$ OPE estimates \cite{Blok:1993va} are shown in the same plot.
The OPE predictions are sensitive to the heavy quark masses.
We take the $\overline{\rm MS}$ mass for the charm quark,
$\bar{m}_c(3\mathrm{~GeV})$ = 1.00~GeV,
and the kinetic mass for the fictitious $b$ quark,
$m_b^{kin}(1\mathrm{~GeV})$ = 2.70(4)~GeV, tuned to reproduce the
$B_s$ meson mass in the simulation using the results of
\cite{Gambino:2017vkx}.
For the OPE matrix elements we employ the results of the semi-leptonic
fit of \cite{Gambino:2016jkc},
although they refer to a light spectator and to the physical $b$ mass.
The dashed lines include $O(1/m^2)$ power corrections,
which are large and tend to improve the agreement with the lattice data
compared to the free quark decay (dotted lines).


\begin{figure}[tb]
  \centering
  \includegraphics[width=8cm]{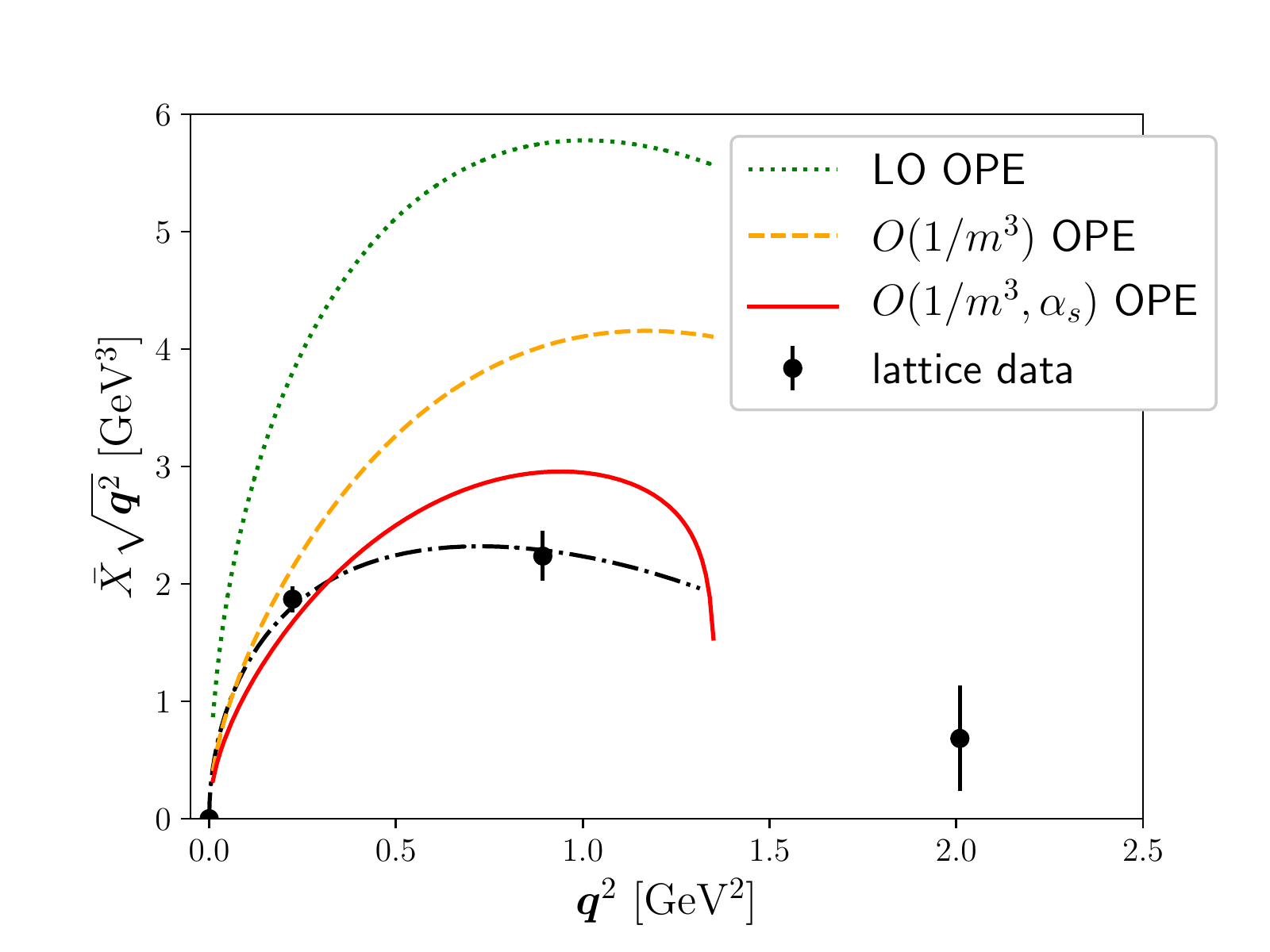}\\
  \caption{
    Integrand of the $\bm{q}^2$-integral plotted in the physical
    unit.
    The dot-dashed curve is an interpolation of the lattice data,
    and the $O(1/m^3, \alpha_s)$ OPE calculation is shown by
    the red curve. 
  }
  \label{fig:integ_q2}
\end{figure}

To obtain the total decay rate, we integrate
$\bar{X}\sqrt{\bm{q}^2}$
over $\bm{q}^2$ as in (\ref{eq:q2integ}).
The vector and axial-vector contributions of different polarizations
are added.
The integrand is shown in Fig.~\ref{fig:integ_q2}.
We fit $\bar{X}^{(l)}/\sqrt{\bm{q}^2}^{2-l}$ by a polynomial of
$\bm{q}^2$ to interpolate the data points.
The fit curve (dot-dashed) is terminated at $\bm{q}^2_{\mathrm{max}}$.
We compare the lattice results with the corresponding OPE prediction
(red curve)
including $O(1/m^3)$ \cite{Gremm:1996df} and
$O(\alpha_s)$ \cite{Aquila:2005hq} terms with $\alpha_s=0.27$.
The power corrections are controlled here by powers of the partonic
energy $\sqrt{m_c^2+\bm{q}^2}$ which ranges between 1 and 1.5~GeV,
significantly less than that for a physical $b$.
They are singular at the partonic endpoint,
where the maximum energy hits the mass-shell of charm quark
and the perturbative corrections show an integrable singularity. 

Integrating the fit to lattice data we obtain 
$\Gamma/|V_{cb}|^2 = 4.9(6)\times 10^{-13}$~GeV,
where only the statistical error is shown.
We note that the total decay rate is about five times smaller than
that of the physical $B_s$ meson, because of the smaller phase
space for the artificially small $b$ quark mass.
On the OPE side, several higher order corrections are available for
the total width, including the complete $O(\alpha_s^2)$
\cite{Pak:2008qt,Melnikov:2008qs}
and the $O(\alpha_s/m^2)$ \cite{Alberti:2013kxa,Mannel:2015jka}
corrections. 
We implement them in the kinetic scheme using the same inputs as above
and obtain
$\Gamma/|V_{cb}|^2 = 5.4(8)\times 10^{-13}$~GeV.
The dominant uncertainty is due to the value of the $b$ quark mass,
but missing higher order corrections and uncertainties on the matrix
elements would also induce an $O(10\%)$ uncertainty. 
Despite these limitations, the agreement between the lattice
and the OPE is remarkable.

An immediate extension of this work is of course the calculation of
the inclusive semi-leptonic decay rate of $B$ mesons and $b$ baryons
($b\to c\ell\bar{\nu}$ and $b\to u\ell\bar{\nu}$).
Moments of kinematical variables, such as the lepton energy moments and
hadronic invariant mass moments, can also be calculated by a slight
modification of the method.
A numerical challenge for the lattice calculation is the large recoil
momentum up to $\sim$ 2.3~GeV, which requires fine lattices to keep the
discretization effects under control.
For $b\to u$ transitions, the experimental analysis involves various
momentum cuts to veto unnecessary $b\to c$ backgrounds.
Our method allows to apply arbitrary kinematical cuts, and a fully
non-perturbative calculation is possible according to the experimental
setup.
A comparison to the OPE calculation at or closer to the physical $b$
mass would provide a valuable test of the OPE, including the
assumption of quark-hadron duality. 
It may also be used to determine the hadronic parameters appearing in
the heavy quark expansion.
The fully non-perturbative lattice calculation can also be applied to
$D$ meson decays, for which the energy release is not sufficiently
large to yield reliable OPE calculations, and where one could observe
the onset of quark-hadron duality. 



The possible applications of the framework are not limited to heavy
quark decays. 
Lepton-nucleon ($\ell N$) scattering is another large area of
application.
Traditionally, it has been analyzed combining perturbation theory and
non-perturbative inputs, such as the parton distribution functions
(PDFs). 
Instead, the method described in this work allows to directly compute
the cross sections without recourse to intermediate quantities like
PDFs, and it opens a new strategy to study the inelastic scatterings.
Moreover, it will make it possible to perform non-perturbative calculation
of low-energy scatterings, which cannot be treated with the
presently available techniques.

\begin{acknowledgments}
  We thank the members of the JLQCD collaboration for discussions and
  for providing the computational framework and lattice data.
  Numerical calculations are performed
  on SX-Aurora TSUBASA at High Energy Accelerator Research
  Organization (KEK) under its Particle, Nuclear and Astro Physics
  Simulation Program, as well as
  on Oakforest-PACS supercomputer operated by Joint Center for Advanced
  High Performance Computing (JCAHPC).
  This work is supported in part by JSPS KAKENHI Grant Number JP26247043
  and by the Post-K and Fugaku supercomputer project through the Joint
  Institute for Computational Fundamental Science (JICFuS). PG 
   is supported in part by the Italian Ministry of Research (MIUR) under grant PRIN 20172LNEEZ.
\end{acknowledgments}

\bibliography{latincl}

\begin{thebibliography}{28}%
\makeatletter
\providecommand \@ifxundefined [1]{%
 \@ifx{#1\undefined}
}%
\providecommand \@ifnum [1]{%
 \ifnum #1\expandafter \@firstoftwo
 \else \expandafter \@secondoftwo
 \fi
}%
\providecommand \@ifx [1]{%
 \ifx #1\expandafter \@firstoftwo
 \else \expandafter \@secondoftwo
 \fi
}%
\providecommand \natexlab [1]{#1}%
\providecommand \enquote  [1]{``#1''}%
\providecommand \bibnamefont  [1]{#1}%
\providecommand \bibfnamefont [1]{#1}%
\providecommand \citenamefont [1]{#1}%
\providecommand \href@noop [0]{\@secondoftwo}%
\providecommand \href [0]{\begingroup \@sanitize@url \@href}%
\providecommand \@href[1]{\@@startlink{#1}\@@href}%
\providecommand \@@href[1]{\endgroup#1\@@endlink}%
\providecommand \@sanitize@url [0]{\catcode `\\12\catcode `\$12\catcode
  `\&12\catcode `\#12\catcode `\^12\catcode `\_12\catcode `\%12\relax}%
\providecommand \@@startlink[1]{}%
\providecommand \@@endlink[0]{}%
\providecommand \url  [0]{\begingroup\@sanitize@url \@url }%
\providecommand \@url [1]{\endgroup\@href {#1}{\urlprefix }}%
\providecommand \urlprefix  [0]{URL }%
\providecommand \Eprint [0]{\href }%
\providecommand \doibase [0]{http://dx.doi.org/}%
\providecommand \selectlanguage [0]{\@gobble}%
\providecommand \bibinfo  [0]{\@secondoftwo}%
\providecommand \bibfield  [0]{\@secondoftwo}%
\providecommand \translation [1]{[#1]}%
\providecommand \BibitemOpen [0]{}%
\providecommand \bibitemStop [0]{}%
\providecommand \bibitemNoStop [0]{.\EOS\space}%
\providecommand \EOS [0]{\spacefactor3000\relax}%
\providecommand \BibitemShut  [1]{\csname bibitem#1\endcsname}%
\let\auto@bib@innerbib\@empty
\bibitem [{\citenamefont {Poggio}\ \emph {et~al.}(1976)\citenamefont {Poggio},
  \citenamefont {Quinn},\ and\ \citenamefont {Weinberg}}]{Poggio:1975af}%
  \BibitemOpen
  \bibfield  {author} {\bibinfo {author} {\bibfnamefont {E.}~\bibnamefont
  {Poggio}}, \bibinfo {author} {\bibfnamefont {H.~R.}\ \bibnamefont {Quinn}}, \
  and\ \bibinfo {author} {\bibfnamefont {S.}~\bibnamefont {Weinberg}},\ }\href
  {\doibase 10.1103/PhysRevD.13.1958} {\bibfield  {journal} {\bibinfo
  {journal} {Phys. Rev. D}\ }\textbf {\bibinfo {volume} {13}},\ \bibinfo
  {pages} {1958} (\bibinfo {year} {1976})}\BibitemShut {NoStop}%
\bibitem [{\citenamefont {Shifman}(2000)}]{Shifman:2000jv}%
  \BibitemOpen
  \bibfield  {author} {\bibinfo {author} {\bibfnamefont {M.~A.}\ \bibnamefont
  {Shifman}},\ }in\ \href {\doibase 10.1142/9789812810458\_0032} {\emph
  {\bibinfo {booktitle} {{8th International Symposium on Heavy Flavor
  Physics}}}},\ Vol.~\bibinfo {volume} {3}\ (\bibinfo  {publisher} {World
  Scientific},\ \bibinfo {address} {Singapore},\ \bibinfo {year} {2000})\ pp.\
  \bibinfo {pages} {1447--1494},\ \Eprint {http://arxiv.org/abs/hep-ph/0009131}
  {arXiv:hep-ph/0009131} \BibitemShut {NoStop}%
\bibitem [{\citenamefont {Bigi}\ and\ \citenamefont
  {Uraltsev}(2001)}]{Bigi:2001ys}%
  \BibitemOpen
  \bibfield  {author} {\bibinfo {author} {\bibfnamefont {I.~I.}\ \bibnamefont
  {Bigi}}\ and\ \bibinfo {author} {\bibfnamefont {N.}~\bibnamefont
  {Uraltsev}},\ }\href {\doibase 10.1142/S0217751X01005535} {\bibfield
  {journal} {\bibinfo  {journal} {Int. J. Mod. Phys. A}\ }\textbf {\bibinfo
  {volume} {16}},\ \bibinfo {pages} {5201} (\bibinfo {year} {2001})},\ \Eprint
  {http://arxiv.org/abs/hep-ph/0106346} {arXiv:hep-ph/0106346} \BibitemShut
  {NoStop}%
\bibitem [{\citenamefont {Alberti}\ \emph {et~al.}(2015)\citenamefont
  {Alberti}, \citenamefont {Gambino}, \citenamefont {Healey},\ and\
  \citenamefont {Nandi}}]{Alberti:2014yda}%
  \BibitemOpen
  \bibfield  {author} {\bibinfo {author} {\bibfnamefont {A.}~\bibnamefont
  {Alberti}}, \bibinfo {author} {\bibfnamefont {P.}~\bibnamefont {Gambino}},
  \bibinfo {author} {\bibfnamefont {K.~J.}\ \bibnamefont {Healey}}, \ and\
  \bibinfo {author} {\bibfnamefont {S.}~\bibnamefont {Nandi}},\ }\href
  {\doibase 10.1103/PhysRevLett.114.061802} {\bibfield  {journal} {\bibinfo
  {journal} {Phys. Rev. Lett.}\ }\textbf {\bibinfo {volume} {114}},\ \bibinfo
  {pages} {061802} (\bibinfo {year} {2015})},\ \Eprint
  {http://arxiv.org/abs/1411.6560} {arXiv:1411.6560 [hep-ph]} \BibitemShut
  {NoStop}%
\bibitem [{\citenamefont {Gambino}\ \emph {et~al.}(2016)\citenamefont
  {Gambino}, \citenamefont {Healey},\ and\ \citenamefont
  {Turczyk}}]{Gambino:2016jkc}%
  \BibitemOpen
  \bibfield  {author} {\bibinfo {author} {\bibfnamefont {P.}~\bibnamefont
  {Gambino}}, \bibinfo {author} {\bibfnamefont {K.~J.}\ \bibnamefont {Healey}},
  \ and\ \bibinfo {author} {\bibfnamefont {S.}~\bibnamefont {Turczyk}},\ }\href
  {\doibase 10.1016/j.physletb.2016.10.023} {\bibfield  {journal} {\bibinfo
  {journal} {Phys. Lett. B}\ }\textbf {\bibinfo {volume} {763}},\ \bibinfo
  {pages} {60} (\bibinfo {year} {2016})},\ \Eprint
  {http://arxiv.org/abs/1606.06174} {arXiv:1606.06174 [hep-ph]} \BibitemShut
  {NoStop}%
\bibitem [{\citenamefont {Tanabashi}\ \emph {et~al.}(2018)\citenamefont
  {Tanabashi} \emph {et~al.}}]{Tanabashi:2018oca}%
  \BibitemOpen
  \bibfield  {author} {\bibinfo {author} {\bibfnamefont {M.}~\bibnamefont
  {Tanabashi}} \emph {et~al.} (\bibinfo {collaboration} {Particle Data
  Group}),\ }\href {\doibase 10.1103/PhysRevD.98.030001} {\bibfield  {journal}
  {\bibinfo  {journal} {Phys. Rev. D}\ }\textbf {\bibinfo {volume} {98}},\
  \bibinfo {pages} {030001} (\bibinfo {year} {2018})}\BibitemShut {NoStop}%
\bibitem [{\citenamefont {Gambino}\ \emph {et~al.}(2019)\citenamefont
  {Gambino}, \citenamefont {Jung},\ and\ \citenamefont
  {Schacht}}]{Gambino:2019sif}%
  \BibitemOpen
  \bibfield  {author} {\bibinfo {author} {\bibfnamefont {P.}~\bibnamefont
  {Gambino}}, \bibinfo {author} {\bibfnamefont {M.}~\bibnamefont {Jung}}, \
  and\ \bibinfo {author} {\bibfnamefont {S.}~\bibnamefont {Schacht}},\ }\href
  {\doibase 10.1016/j.physletb.2019.06.039} {\bibfield  {journal} {\bibinfo
  {journal} {Phys. Lett. B}\ }\textbf {\bibinfo {volume} {795}},\ \bibinfo
  {pages} {386} (\bibinfo {year} {2019})},\ \Eprint
  {http://arxiv.org/abs/1905.08209} {arXiv:1905.08209 [hep-ph]} \BibitemShut
  {NoStop}%
\bibitem [{\citenamefont {Aoki}\ \emph {et~al.}(2020)\citenamefont {Aoki} \emph
  {et~al.}}]{Aoki:2019cca}%
  \BibitemOpen
  \bibfield  {author} {\bibinfo {author} {\bibfnamefont {S.}~\bibnamefont
  {Aoki}} \emph {et~al.} (\bibinfo {collaboration} {Flavour Lattice Averaging
  Group}),\ }\href {\doibase 10.1140/epjc/s10052-019-7354-7} {\bibfield
  {journal} {\bibinfo  {journal} {Eur. Phys. J. C}\ }\textbf {\bibinfo {volume}
  {80}},\ \bibinfo {pages} {113} (\bibinfo {year} {2020})},\ \Eprint
  {http://arxiv.org/abs/1902.08191} {arXiv:1902.08191 [hep-lat]} \BibitemShut
  {NoStop}%
\bibitem [{\citenamefont {Hashimoto}(2017)}]{Hashimoto:2017wqo}%
  \BibitemOpen
  \bibfield  {author} {\bibinfo {author} {\bibfnamefont {S.}~\bibnamefont
  {Hashimoto}},\ }\href {\doibase 10.1093/ptep/ptx052} {\bibfield  {journal}
  {\bibinfo  {journal} {PTEP}\ }\textbf {\bibinfo {volume} {2017}},\ \bibinfo
  {pages} {053B03} (\bibinfo {year} {2017})},\ \Eprint
  {http://arxiv.org/abs/1703.01881} {arXiv:1703.01881 [hep-lat]} \BibitemShut
  {NoStop}%
\bibitem [{\citenamefont {Hansen}\ \emph {et~al.}(2017)\citenamefont {Hansen},
  \citenamefont {Meyer},\ and\ \citenamefont {Robaina}}]{Hansen:2017mnd}%
  \BibitemOpen
  \bibfield  {author} {\bibinfo {author} {\bibfnamefont {M.~T.}\ \bibnamefont
  {Hansen}}, \bibinfo {author} {\bibfnamefont {H.~B.}\ \bibnamefont {Meyer}}, \
  and\ \bibinfo {author} {\bibfnamefont {D.}~\bibnamefont {Robaina}},\ }\href
  {\doibase 10.1103/PhysRevD.96.094513} {\bibfield  {journal} {\bibinfo
  {journal} {Phys. Rev. D}\ }\textbf {\bibinfo {volume} {96}},\ \bibinfo
  {pages} {094513} (\bibinfo {year} {2017})},\ \Eprint
  {http://arxiv.org/abs/1704.08993} {arXiv:1704.08993 [hep-lat]} \BibitemShut
  {NoStop}%
\bibitem [{\citenamefont {Bernecker}\ and\ \citenamefont
  {Meyer}(2011)}]{Bernecker:2011gh}%
  \BibitemOpen
  \bibfield  {author} {\bibinfo {author} {\bibfnamefont {D.}~\bibnamefont
  {Bernecker}}\ and\ \bibinfo {author} {\bibfnamefont {H.~B.}\ \bibnamefont
  {Meyer}},\ }\href {\doibase 10.1140/epja/i2011-11148-6} {\bibfield  {journal}
  {\bibinfo  {journal} {Eur. Phys. J. A}\ }\textbf {\bibinfo {volume} {47}},\
  \bibinfo {pages} {148} (\bibinfo {year} {2011})},\ \Eprint
  {http://arxiv.org/abs/1107.4388} {arXiv:1107.4388 [hep-lat]} \BibitemShut
  {NoStop}%
\bibitem [{\citenamefont {Feng}\ \emph {et~al.}(2013)\citenamefont {Feng},
  \citenamefont {Hashimoto}, \citenamefont {Hotzel}, \citenamefont {Jansen},
  \citenamefont {Petschlies},\ and\ \citenamefont {Renner}}]{Feng:2013xsa}%
  \BibitemOpen
  \bibfield  {author} {\bibinfo {author} {\bibfnamefont {X.}~\bibnamefont
  {Feng}}, \bibinfo {author} {\bibfnamefont {S.}~\bibnamefont {Hashimoto}},
  \bibinfo {author} {\bibfnamefont {G.}~\bibnamefont {Hotzel}}, \bibinfo
  {author} {\bibfnamefont {K.}~\bibnamefont {Jansen}}, \bibinfo {author}
  {\bibfnamefont {M.}~\bibnamefont {Petschlies}}, \ and\ \bibinfo {author}
  {\bibfnamefont {D.~B.}\ \bibnamefont {Renner}},\ }\href {\doibase
  10.1103/PhysRevD.88.034505} {\bibfield  {journal} {\bibinfo  {journal} {Phys.
  Rev. D}\ }\textbf {\bibinfo {volume} {88}},\ \bibinfo {pages} {034505}
  (\bibinfo {year} {2013})},\ \Eprint {http://arxiv.org/abs/1305.5878}
  {arXiv:1305.5878 [hep-lat]} \BibitemShut {NoStop}%
\bibitem [{\citenamefont {Francis}\ \emph {et~al.}(2013)\citenamefont
  {Francis}, \citenamefont {Jaeger}, \citenamefont {Meyer},\ and\ \citenamefont
  {Wittig}}]{Francis:2013fzp}%
  \BibitemOpen
  \bibfield  {author} {\bibinfo {author} {\bibfnamefont {A.}~\bibnamefont
  {Francis}}, \bibinfo {author} {\bibfnamefont {B.}~\bibnamefont {Jaeger}},
  \bibinfo {author} {\bibfnamefont {H.~B.}\ \bibnamefont {Meyer}}, \ and\
  \bibinfo {author} {\bibfnamefont {H.}~\bibnamefont {Wittig}},\ }\href
  {\doibase 10.1103/PhysRevD.88.054502} {\bibfield  {journal} {\bibinfo
  {journal} {Phys. Rev. D}\ }\textbf {\bibinfo {volume} {88}},\ \bibinfo
  {pages} {054502} (\bibinfo {year} {2013})},\ \Eprint
  {http://arxiv.org/abs/1306.2532} {arXiv:1306.2532 [hep-lat]} \BibitemShut
  {NoStop}%
\bibitem [{\citenamefont {Lehner}\ and\ \citenamefont
  {Meyer}(2020)}]{Lehner:2020crt}%
  \BibitemOpen
  \bibfield  {author} {\bibinfo {author} {\bibfnamefont {C.}~\bibnamefont
  {Lehner}}\ and\ \bibinfo {author} {\bibfnamefont {A.~S.}\ \bibnamefont
  {Meyer}},\ }\href {\doibase 10.1103/PhysRevD.101.074515} {\bibfield
  {journal} {\bibinfo  {journal} {Phys. Rev. D}\ }\textbf {\bibinfo {volume}
  {101}},\ \bibinfo {pages} {074515} (\bibinfo {year} {2020})},\ \Eprint
  {http://arxiv.org/abs/2003.04177} {arXiv:2003.04177 [hep-lat]} \BibitemShut
  {NoStop}%
\bibitem [{\citenamefont {Tomii}\ \emph {et~al.}(2017)\citenamefont {Tomii},
  \citenamefont {Cossu}, \citenamefont {Fahy}, \citenamefont {Fukaya},
  \citenamefont {Hashimoto}, \citenamefont {Kaneko},\ and\ \citenamefont
  {Noaki}}]{Tomii:2017cbt}%
  \BibitemOpen
  \bibfield  {author} {\bibinfo {author} {\bibfnamefont {M.}~\bibnamefont
  {Tomii}}, \bibinfo {author} {\bibfnamefont {G.}~\bibnamefont {Cossu}},
  \bibinfo {author} {\bibfnamefont {B.}~\bibnamefont {Fahy}}, \bibinfo {author}
  {\bibfnamefont {H.}~\bibnamefont {Fukaya}}, \bibinfo {author} {\bibfnamefont
  {S.}~\bibnamefont {Hashimoto}}, \bibinfo {author} {\bibfnamefont
  {T.}~\bibnamefont {Kaneko}}, \ and\ \bibinfo {author} {\bibfnamefont
  {J.}~\bibnamefont {Noaki}} (\bibinfo {collaboration} {JLQCD}),\ }\href
  {\doibase 10.1103/PhysRevD.96.054511} {\bibfield  {journal} {\bibinfo
  {journal} {Phys. Rev. D}\ }\textbf {\bibinfo {volume} {96}},\ \bibinfo
  {pages} {054511} (\bibinfo {year} {2017})},\ \Eprint
  {http://arxiv.org/abs/1703.06249} {arXiv:1703.06249 [hep-lat]} \BibitemShut
  {NoStop}%
\bibitem [{\citenamefont {Boyle}\ \emph {et~al.}(2018)\citenamefont {Boyle},
  \citenamefont {Hudspith}, \citenamefont {Izubuchi}, \citenamefont {Jüttner},
  \citenamefont {Lehner}, \citenamefont {Lewis}, \citenamefont {Maltman},
  \citenamefont {Ohki}, \citenamefont {Portelli},\ and\ \citenamefont
  {Spraggs}}]{Boyle:2018ilm}%
  \BibitemOpen
  \bibfield  {author} {\bibinfo {author} {\bibfnamefont {P.}~\bibnamefont
  {Boyle}}, \bibinfo {author} {\bibfnamefont {R.~J.}\ \bibnamefont {Hudspith}},
  \bibinfo {author} {\bibfnamefont {T.}~\bibnamefont {Izubuchi}}, \bibinfo
  {author} {\bibfnamefont {A.}~\bibnamefont {Jüttner}}, \bibinfo {author}
  {\bibfnamefont {C.}~\bibnamefont {Lehner}}, \bibinfo {author} {\bibfnamefont
  {R.}~\bibnamefont {Lewis}}, \bibinfo {author} {\bibfnamefont
  {K.}~\bibnamefont {Maltman}}, \bibinfo {author} {\bibfnamefont
  {H.}~\bibnamefont {Ohki}}, \bibinfo {author} {\bibfnamefont {A.}~\bibnamefont
  {Portelli}}, \ and\ \bibinfo {author} {\bibfnamefont {M.}~\bibnamefont
  {Spraggs}} (\bibinfo {collaboration} {RBC, UKQCD}),\ }\href {\doibase
  10.1103/PhysRevLett.121.202003} {\bibfield  {journal} {\bibinfo  {journal}
  {Phys. Rev. Lett.}\ }\textbf {\bibinfo {volume} {121}},\ \bibinfo {pages}
  {202003} (\bibinfo {year} {2018})},\ \Eprint
  {http://arxiv.org/abs/1803.07228} {arXiv:1803.07228 [hep-lat]} \BibitemShut
  {NoStop}%
\bibitem [{\citenamefont {Bailas}\ \emph {et~al.}(2020)\citenamefont {Bailas},
  \citenamefont {Hashimoto},\ and\ \citenamefont {Ishikawa}}]{Bailas:2020qmv}%
  \BibitemOpen
  \bibfield  {author} {\bibinfo {author} {\bibfnamefont {G.}~\bibnamefont
  {Bailas}}, \bibinfo {author} {\bibfnamefont {S.}~\bibnamefont {Hashimoto}}, \
  and\ \bibinfo {author} {\bibfnamefont {T.}~\bibnamefont {Ishikawa}},\ }\href
  {\doibase 10.1093/ptep/ptaa044} {\bibfield  {journal} {\bibinfo  {journal}
  {PTEP}\ }\textbf {\bibinfo {volume} {2020}},\ \bibinfo {pages} {043B07}
  (\bibinfo {year} {2020})},\ \Eprint {http://arxiv.org/abs/2001.11779}
  {arXiv:2001.11779 [hep-lat]} \BibitemShut {NoStop}%
\bibitem [{\citenamefont {Bulava}\ and\ \citenamefont
  {Hansen}(2019)}]{Bulava:2019kbi}%
  \BibitemOpen
  \bibfield  {author} {\bibinfo {author} {\bibfnamefont {J.}~\bibnamefont
  {Bulava}}\ and\ \bibinfo {author} {\bibfnamefont {M.~T.}\ \bibnamefont
  {Hansen}},\ }\href {\doibase 10.1103/PhysRevD.100.034521} {\bibfield
  {journal} {\bibinfo  {journal} {Phys. Rev. D}\ }\textbf {\bibinfo {volume}
  {100}},\ \bibinfo {pages} {034521} (\bibinfo {year} {2019})},\ \Eprint
  {http://arxiv.org/abs/1903.11735} {arXiv:1903.11735 [hep-lat]} \BibitemShut
  {NoStop}%
\bibitem [{\citenamefont {Manohar}\ and\ \citenamefont
  {Wise}(1994)}]{Manohar:1993qn}%
  \BibitemOpen
  \bibfield  {author} {\bibinfo {author} {\bibfnamefont {A.~V.}\ \bibnamefont
  {Manohar}}\ and\ \bibinfo {author} {\bibfnamefont {M.~B.}\ \bibnamefont
  {Wise}},\ }\href {\doibase 10.1103/PhysRevD.49.1310} {\bibfield  {journal}
  {\bibinfo  {journal} {Phys. Rev. D}\ }\textbf {\bibinfo {volume} {49}},\
  \bibinfo {pages} {1310} (\bibinfo {year} {1994})},\ \Eprint
  {http://arxiv.org/abs/hep-ph/9308246} {arXiv:hep-ph/9308246} \BibitemShut
  {NoStop}%
\bibitem [{\citenamefont {Blok}\ \emph {et~al.}(1994)\citenamefont {Blok},
  \citenamefont {Koyrakh}, \citenamefont {Shifman},\ and\ \citenamefont
  {Vainshtein}}]{Blok:1993va}%
  \BibitemOpen
  \bibfield  {author} {\bibinfo {author} {\bibfnamefont {B.}~\bibnamefont
  {Blok}}, \bibinfo {author} {\bibfnamefont {L.}~\bibnamefont {Koyrakh}},
  \bibinfo {author} {\bibfnamefont {M.~A.}\ \bibnamefont {Shifman}}, \ and\
  \bibinfo {author} {\bibfnamefont {A.}~\bibnamefont {Vainshtein}},\ }\href
  {\doibase 10.1103/PhysRevD.50.3572} {\bibfield  {journal} {\bibinfo
  {journal} {Phys. Rev. D}\ }\textbf {\bibinfo {volume} {49}},\ \bibinfo
  {pages} {3356} (\bibinfo {year} {1994})},\ \bibinfo {note} {[Erratum: Phys.
  Rev. D 50, 3572 (1994)]},\ \Eprint {http://arxiv.org/abs/hep-ph/9307247}
  {arXiv:hep-ph/9307247} \BibitemShut {NoStop}%
\bibitem [{\citenamefont {Nakayama}\ \emph {et~al.}(2016)\citenamefont
  {Nakayama}, \citenamefont {Fahy},\ and\ \citenamefont
  {Hashimoto}}]{Nakayama:2016atf}%
  \BibitemOpen
  \bibfield  {author} {\bibinfo {author} {\bibfnamefont {K.}~\bibnamefont
  {Nakayama}}, \bibinfo {author} {\bibfnamefont {B.}~\bibnamefont {Fahy}}, \
  and\ \bibinfo {author} {\bibfnamefont {S.}~\bibnamefont {Hashimoto}},\ }\href
  {\doibase 10.1103/PhysRevD.94.054507} {\bibfield  {journal} {\bibinfo
  {journal} {Phys. Rev. D}\ }\textbf {\bibinfo {volume} {94}},\ \bibinfo
  {pages} {054507} (\bibinfo {year} {2016})},\ \Eprint
  {http://arxiv.org/abs/1606.01002} {arXiv:1606.01002 [hep-lat]} \BibitemShut
  {NoStop}%
\bibitem [{\citenamefont {Gambino}\ \emph {et~al.}(2017)\citenamefont
  {Gambino}, \citenamefont {Melis},\ and\ \citenamefont
  {Simula}}]{Gambino:2017vkx}%
  \BibitemOpen
  \bibfield  {author} {\bibinfo {author} {\bibfnamefont {P.}~\bibnamefont
  {Gambino}}, \bibinfo {author} {\bibfnamefont {A.}~\bibnamefont {Melis}}, \
  and\ \bibinfo {author} {\bibfnamefont {S.}~\bibnamefont {Simula}},\ }\href
  {\doibase 10.1103/PhysRevD.96.014511} {\bibfield  {journal} {\bibinfo
  {journal} {Phys. Rev. D}\ }\textbf {\bibinfo {volume} {96}},\ \bibinfo
  {pages} {014511} (\bibinfo {year} {2017})},\ \Eprint
  {http://arxiv.org/abs/1704.06105} {arXiv:1704.06105 [hep-lat]} \BibitemShut
  {NoStop}%
\bibitem [{\citenamefont {Gremm}\ and\ \citenamefont
  {Kapustin}(1997)}]{Gremm:1996df}%
  \BibitemOpen
  \bibfield  {author} {\bibinfo {author} {\bibfnamefont {M.}~\bibnamefont
  {Gremm}}\ and\ \bibinfo {author} {\bibfnamefont {A.}~\bibnamefont
  {Kapustin}},\ }\href {\doibase 10.1103/PhysRevD.55.6924} {\bibfield
  {journal} {\bibinfo  {journal} {Phys. Rev. D}\ }\textbf {\bibinfo {volume}
  {55}},\ \bibinfo {pages} {6924} (\bibinfo {year} {1997})},\ \Eprint
  {http://arxiv.org/abs/hep-ph/9603448} {arXiv:hep-ph/9603448} \BibitemShut
  {NoStop}%
\bibitem [{\citenamefont {Aquila}\ \emph {et~al.}(2005)\citenamefont {Aquila},
  \citenamefont {Gambino}, \citenamefont {Ridolfi},\ and\ \citenamefont
  {Uraltsev}}]{Aquila:2005hq}%
  \BibitemOpen
  \bibfield  {author} {\bibinfo {author} {\bibfnamefont {V.}~\bibnamefont
  {Aquila}}, \bibinfo {author} {\bibfnamefont {P.}~\bibnamefont {Gambino}},
  \bibinfo {author} {\bibfnamefont {G.}~\bibnamefont {Ridolfi}}, \ and\
  \bibinfo {author} {\bibfnamefont {N.}~\bibnamefont {Uraltsev}},\ }\href
  {\doibase 10.1016/j.nuclphysb.2005.04.031} {\bibfield  {journal} {\bibinfo
  {journal} {Nucl. Phys. B}\ }\textbf {\bibinfo {volume} {719}},\ \bibinfo
  {pages} {77} (\bibinfo {year} {2005})},\ \Eprint
  {http://arxiv.org/abs/hep-ph/0503083} {arXiv:hep-ph/0503083} \BibitemShut
  {NoStop}%
\bibitem [{\citenamefont {Pak}\ and\ \citenamefont
  {Czarnecki}(2008)}]{Pak:2008qt}%
  \BibitemOpen
  \bibfield  {author} {\bibinfo {author} {\bibfnamefont {A.}~\bibnamefont
  {Pak}}\ and\ \bibinfo {author} {\bibfnamefont {A.}~\bibnamefont
  {Czarnecki}},\ }\href {\doibase 10.1103/PhysRevLett.100.241807} {\bibfield
  {journal} {\bibinfo  {journal} {Phys. Rev. Lett.}\ }\textbf {\bibinfo
  {volume} {100}},\ \bibinfo {pages} {241807} (\bibinfo {year} {2008})},\
  \Eprint {http://arxiv.org/abs/0803.0960} {arXiv:0803.0960 [hep-ph]}
  \BibitemShut {NoStop}%
\bibitem [{\citenamefont {Melnikov}(2008)}]{Melnikov:2008qs}%
  \BibitemOpen
  \bibfield  {author} {\bibinfo {author} {\bibfnamefont {K.}~\bibnamefont
  {Melnikov}},\ }\href {\doibase 10.1016/j.physletb.2008.07.089} {\bibfield
  {journal} {\bibinfo  {journal} {Phys. Lett. B}\ }\textbf {\bibinfo {volume}
  {666}},\ \bibinfo {pages} {336} (\bibinfo {year} {2008})},\ \Eprint
  {http://arxiv.org/abs/0803.0951} {arXiv:0803.0951 [hep-ph]} \BibitemShut
  {NoStop}%
\bibitem [{\citenamefont {Alberti}\ \emph {et~al.}(2014)\citenamefont
  {Alberti}, \citenamefont {Gambino},\ and\ \citenamefont
  {Nandi}}]{Alberti:2013kxa}%
  \BibitemOpen
  \bibfield  {author} {\bibinfo {author} {\bibfnamefont {A.}~\bibnamefont
  {Alberti}}, \bibinfo {author} {\bibfnamefont {P.}~\bibnamefont {Gambino}}, \
  and\ \bibinfo {author} {\bibfnamefont {S.}~\bibnamefont {Nandi}},\ }\href
  {\doibase 10.1007/JHEP01(2014)147} {\bibfield  {journal} {\bibinfo  {journal}
  {JHEP}\ }\textbf {\bibinfo {volume} {01}},\ \bibinfo {pages} {147} (\bibinfo
  {year} {2014})},\ \Eprint {http://arxiv.org/abs/1311.7381} {arXiv:1311.7381
  [hep-ph]} \BibitemShut {NoStop}%
\bibitem [{\citenamefont {Mannel}\ \emph {et~al.}(2015)\citenamefont {Mannel},
  \citenamefont {Pivovarov},\ and\ \citenamefont {Rosenthal}}]{Mannel:2015jka}%
  \BibitemOpen
  \bibfield  {author} {\bibinfo {author} {\bibfnamefont {T.}~\bibnamefont
  {Mannel}}, \bibinfo {author} {\bibfnamefont {A.~A.}\ \bibnamefont
  {Pivovarov}}, \ and\ \bibinfo {author} {\bibfnamefont {D.}~\bibnamefont
  {Rosenthal}},\ }\href {\doibase 10.1103/PhysRevD.92.054025} {\bibfield
  {journal} {\bibinfo  {journal} {Phys. Rev. D}\ }\textbf {\bibinfo {volume}
  {92}},\ \bibinfo {pages} {054025} (\bibinfo {year} {2015})},\ \Eprint
  {http://arxiv.org/abs/1506.08167} {arXiv:1506.08167 [hep-ph]} \BibitemShut
  {NoStop}%
\end{thebibliography}%
\end{document}